\documentclass[oribibl]{llncs}
\usepackage{llncsdoc}
\usepackage{amssymb,amsmath}

\usepackage{graphics}
\newtheorem{thm}{Theorem}
\newtheorem{clm}{Claim}

\def \no {\noindent}

\begin{document}
\title{New Polynomial Case for Efficient Domination in $P_6$-free Graphs}

\author{T. Karthick}
\institute{Computer Science Unit,\\ Indian Statistical Institute,
Chennai Centre,\\
Chennai-600113, India.\\  \email{karthick@isichennai.res.in} }

\titlerunning{Efficient domination in graph classes}
\maketitle

\begin{abstract}
In a graph $G$, an {\it efficient dominating set} is a subset $D$ of vertices  such that $D$ is an
independent set and each vertex outside $D$ has exactly one neighbor in $D$.
The {\textsc{Efficient Dominating Set}} problem  (EDS) asks for the existence of an efficient dominating set in a given graph $G$.
The EDS is known to be $NP$-complete for $P_7$-free graphs, and is known to be polynomial time solvable for $P_5$-free graphs.
However, the computational complexity of the EDS problem  is unknown for $P_6$-free graphs.
In this paper, we show that the EDS problem  can be solved in polynomial time for a subclass of $P_6$-free graphs, namely
 ($P_6$, banner)-free graphs.
\end{abstract}


\section{Introduction}
Throughout this paper, let $G = (V, E)$ be a finite, undirected and simple graph. We follow West \cite{West} for standard notations and terminology.
If $\cal{F}$ is a family of graphs,
  a graph $G$ is said to be {\it $\cal{F}$-free} if it contains no
induced subgraph isomorphic to any graph in $\cal{F}$. Let $P_t$ denotes the path on $t$ vertices.

In a graph $G$, a subset $D \subseteq V$ is a {\it dominating set} if each vertex outside $D$ has some neighbor in $D$. An
{\it efficient dominating set} is a dominating set $D$ such that $D$ is an independent set and each vertex outside $D$ has exactly one neighbor in $D$.
Efficient dominating sets were introduced by Biggs \cite{Biggs}, and are also called perfect codes, perfect dominating sets and independent perfect dominating sets in the literature. We refer to \cite{HHS} for more information on efficient domination in graphs.
The {\textsc{Efficient Dominating Set}} problem  (EDS) asks for the existence of an efficient dominating set in a given graph $G$. 
The EDS problem is motivated by various applications such as coding theory and resource allocation in parallel computer networks; see \cite{Biggs, LS}.

The EDS is known to be $NP$-complete in general, and is known to be $NP$-complete for several restricted classes of graphs chordal bipartite graphs \cite{LT-2}, planar bipartite graphs \cite{LT-2}, and planar graphs with maximum degree three \cite{FH}. However, the EDS is solvable in polynomial time  for split graphs \cite{CL-1}, cocomparability graphs \cite{CPC}, interval graphs \cite{CL-2}, circular-arc graphs \cite{CL-2}, and for many more classes of graphs  (see \cite{BMN} and the references therein).
In particular,  EDS is $NP$-complete  for $2P_3$-free chordal graphs \cite{SS}, and hence EDS remains $NP$-complete  for $P_7$-free graphs.
 Milanic \cite{M} showed that the EDS is solvable in polynomial time for $P_5$-free graphs. Brandst\"adt and Le \cite{BL} showed that the EDS is solvable in polynomial time for (E, xNet)-free graphs, thereby extending the result on $P_5$-free graphs. However, the computational complexity of EDS is unknown for $P_6$-free graphs. In \cite{BMN}, Brandst\"adt et al.  showed that the EDS is solvable in polynomial time for ($P_6, S_{1, 2, 2}$)-free graphs.  Recently, the author showed that EDS is solvable in polynomial time for ($P_6, S_{1, 1, 3}$)-free graphs, and ($P_6$, bull)-free graphs \cite{K2}. We refer to Figure 1 of \cite{BMN} for the recent complexity status of EDS on several graph classes. 
 
 In this paper, we show that the EDS problem  can be solved in polynomial time for another subclass of $P_6$-free graphs, namely ($P_6$, banner)-free graphs, where a {\it banner} is the graph obtained from a  chordless cycle on four vertices by adding a vertex that has exactly one neighbor on
the cycle. A banner is also called as $P$, $4$-apple and $A_4$ in the literature.

The class of banner-free graphs includes several well studied classes of graphs in the literature such as: $P_4$-free graphs (or co-graphs), $K_{1, 3}$-free graphs (or claw-free graphs), and $C_4$-free graphs. Note that from the $NP$-completeness result for $K_{1, 3}$-free graphs \cite{LT-1}, it follows that for banner-free graphs, the EDS remains $NP$-complete.

 If $G$ is a graph, and if $S \subseteq V(G)$, then $G[S]$ denote the subgraph induced by $S$ in $G$.
  For any two vertices $u$ and $v$ in $G$, dist$_G(u, v)$ denote the {\it distance} between $u$ and $v$ in $G$. The {\it square} of  a graph $G = (V, E)$ is the graph
$G^2 = (V, E^2)$ such that $uv \in E^2$ if and only if $\mbox{dist}_G(u, v) \in \{1, 2\}$.

The following lemma given in \cite{M} (see also \cite{BL}) relates the EDS problem on $G$ and the {\textsc{Maximum Weight Independent Set}} (MWIS) problem on $G^2$.

\begin{lemma} \cite{M} \label{EDS-MWIS-reduction}
Let $G$ be a graph with vertex weight $w(v)$ equal to the number of neighbors of $v$ plus one.
Then the following statements are equivalent for any subset $D \subseteq V$:
\begin{enumerate}\item[(i)] $D$ is an efficient dominating set in $G$.
\item[(ii)] $D$ is a minimum weight dominating set in $G$ with $\sum_{v\in D} w(v) = |V|$.
\item[(iii)] $D$ is a maximum weight independent set in $G^2$ with $\sum_{v\in D} w(v) = |V|$. \hfill{$\Box$}
\end{enumerate}
\end{lemma}

Thus, the EDS problem on a graph class $\cal{G}$ can be reduced to the MWIS problem on the squares of graphs in $\cal{G}$.

In Section 2, we show that if $G$ is a ($P_6$, banner)-free graph that has an efficient dominating set, then $G^2$ is also ($P_6$, banner)-free  (Theorems 1 and 2). Since MWIS can be solved in polynomial time for ($P_6$, banner)-free graphs \cite{BKLM, K}, we deduce that the EDS problem  can be solved in polynomial time for ($P_6$, banner)-free graphs, by Lemma \ref{EDS-MWIS-reduction} (Theorem 3).

\section{EDS in ($P_6$, banner)-free graphs}

In this section, we show that the EDS can be solved efficiently in ($P_6$, banner)-free graphs. First, we prove the following:

\begin{thm}\label{P6-banner-free-ed-implies-P6-free}
Let $G = (V, E)$ be a ($P_6$, banner)-free graph. If $G$ has an efficient dominating set, then $G^2$ is $P_6$-free.
\end{thm}

\no{\it Proof of Theorem \ref{P6-banner-free-ed-implies-P6-free}}: Let $G$ be a ($P_6$, banner)-free graph having an efficient dominating set $D$, and assume to the contrary that $G^2$ contains an induced
$P_6$, say with vertices $v_1, v_2, v_3, v_4, v_5$ and $v_6$, and edges $v_1v_2, v_2v_3, v_3v_4, v_4v_5,$ and $v_5v_6$.  Then $\mbox{dist}_G(v_1, v_2) \leq 2, \mbox{dist}_G(v_2, v_3) \leq 2, \mbox{dist}_G(v_3, v_4) \leq 2$, $\mbox{dist}_G(v_4, v_5) \leq 2$, and $\mbox{dist}_G(v_5, v_6) \leq 2$ while $\mbox{dist}_G(v_1, v_3) \geq 3, \mbox{dist}_G(v_1, v_4) \geq 3, \mbox{dist}_G(v_1, v_5) \geq 3, \mbox{dist}_G(v_1, v_6) \geq 3, \mbox{dist}_G(v_2, v_4) \geq 3, ~~~~\mbox{dist}_G(v_2, v_5) \geq 3, ~~~~\mbox{dist}_G(v_2, v_6) \geq 3, ~~~~$ $\mbox{dist}_G(v_3, v_5) \geq 3, \mbox{dist}_G(v_3, v_6)$ $ \geq 3$, and $\mbox{dist}_G(v_4,$ $ v_6) \geq 3$. We often use these distance properties implicitly in the remaining proof.

\vspace{0.25cm}
\noindent{\bf Case 1}: Suppose that $\mbox{dist}_G(v_5, v_6) = 1$.

Since $\mbox{dist}_G(v_4, v_6) \geq 3$, we have $\mbox{dist}_G(v_4, v_5) = 2$, and so there exists $d \in V$ such that $dv_4, dv_5 \in E$. If $\mbox{dist}_G(v_3, v_4) = 1$, then since $\mbox{dist}_G(v_2, v_4) \geq 3$, we have $\mbox{dist}_G(v_2, v_3) = 2$, and hence there exists $x \in V$ such that $xv_2, xv_3 \in E$. Now, (i) if $xd \in E$, then $\{v_4, v_3, x, d, v_5\}$ will induce a banner in $G$, and (ii) if $xd \notin E$ then $\{v_6, v_5, d, v_4, v_3, x\}$ will induce a $P_6$ in $G$,  a contradiction.

 So, assume that $\mbox{dist}_G(v_3, v_4) =2$, and hence there exists $c \in V$ such that $cv_3, cv_4 \in E$.  Then $cd \in E$ (otherwise, $G[\{v_6, v_5, d, v_4, c, v_3\}]$ is a $P_6$ in $G$). So, $\mbox{dist}_G(v_2, v_3) = 2$ (otherwise, $G[\{v_6, v_5, d, c, v_3, v_2\}]$ is a $P_6$ in $G$), and hence there exists $b \in V$ such that $bv_2, bv_3 \in E$. Then $bc \in E$ (otherwise, since $G[\{v_6, v_5, d, c, v_3, b\}]$ is not an induced $P_6$ in $G$, we have $bd \in E$. But, then $G[\{b, v_3, c, d, v_5\}]$ is a banner in $G$, a contradiction), and hence $bd \in E$ (otherwise, $G[\{v_6, v_5, d, c, b, v_2\}]$ is a $P_6$ in $G$). Then  $\mbox{dist}_G(v_1, v_2) = 2$ (otherwise, $G[\{v_6, v_5, d, b, v_2, v_1\}]$ is a $P_6$ in $G$), and hence there exists $a \in V$ such that $av_1, av_2 \in E$. Then $ab \in E$ (otherwise, $\{a, v_2, b, d, v_5, v_6\}$ will induce a banner or $P_6$ in $G$  according as $ad \in E$ or $ad \notin E$ respectively, a contradiction), and hence $ad \in E$ (otherwise, $G[\{v_1, a, b, d, v_5, v_6\}]$ is a $P_6$ in $G$).

  Now, we have the following:

\begin{clm}\label{a-b-v2-notin-D} $a, v_2, b \notin D$.
\end{clm}
\no{\it Proof of Claim \ref{a-b-v2-notin-D}} : We prove the claim by assuming the contrary one by one as follows:
\begin{enumerate}

\item[(i)] On the contrary, assume that $a \in D$. Then by the definition of $D$, we have $b, v_3, c \notin D$. So, there exists $v_3' \in D$ such that $v_3v_3' \in E$. Then by using the definition of $D$ and by the distance properties, we see that $G[\{v_3', v_3, b, d, v_5, v_6\}]$ is a $P_6$ in $G$, a contradiction.  Hence, $a \notin D$.

\item[(ii)]  On the contrary, assume that $v_2 \in D$ (or $b \in D$). Then by the definition of $D$, we have $a, v_1 \notin D$.  So, there exists $v_1' \in D$ such that $v_1v_1' \in E$. Then by using the definition of $D$ and by the distance properties, we see that $\{v_1', v_1, a, d, v_5, v_6\}$ will induce a  $P_6$ or banner in $G$, a contradiction.  Hence, $v_2, b \notin D$.  $\blacklozenge$

\end{enumerate}

Since $a, v_2, b \notin D$ (by Claim \ref{a-b-v2-notin-D}), there exists $v_2' \in D$ such that $v_2v_2' \in E$.  Then $v_2'a, v_2'b \in E$ (otherwise,
$\{v_2', v_2, a, b, d, v_5, v_6\}$ will induce a banner or a $P_6$ in $G$).

So, $v_1 \notin D$, and hence there exists $v_1' \in D$ such that $v_1v_1' \in E$. Then we show the following:

\begin{clm}\label{v1'-neq-v2'} $v_1' \neq v_2'$ (that is, $v_1v_2' \notin E$).
\end{clm}
\no{\it Proof of Claim \ref{v1'-neq-v2'}} : Assume the contrary. Then $v_2'd \in E$ (otherwise, $G[\{v_1, v_2', b, d,$ $ v_5, v_6\}]$ is a $P_6$ in $G$). Then since $b, v_3, c \notin D$ (by the definition of $D$), there exists $v_3' \in D$ such that $v_3v_3' \in E$. Also, since $\mbox{dist}_G(v_1, v_3) \geq 3$, we have $v_2' \neq v_3'$. Then by using the definition of $D$ and by the distance properties, we see that $G[\{v_3', v_3, b, d, v_5, v_6\}]$ is a $P_6$ in $G$, a contradiction. $\blacklozenge$

Now, if $v_1'd \in E$, then $\{v_1', v_1, a, d, v_5\}$ will induce a banner in $G$, and if $v_1'd \notin E$, then $\{v_1', v_1, a, d, v_5, v_6\}$ will induce a $P_6$ in $G$, a contradiction.

\vspace{0.25cm}
\noindent{\bf Case 2}: Suppose that $\mbox{dist}_G(v_1, v_2) = 2 = \mbox{dist}_G(v_5, v_6)$.

Then there exist $a, e \in V$ such that $av_1, av_2, ev_5, ev_6 \in E$. If  $\mbox{dist}_G(v_2, v_3)=1$, then since $\mbox{dist}_G(v_2, v_4) \geq 3$,  we have $\mbox{dist}_G(v_3, v_4) = 2$, and hence there exists $x \in V$ such that $xv_3, xv_4 \in E$. But, then $\{v_1, a, v_2, v_3, x, v_4\}$ will induce either  a banner or a $P_6$ in $G$, according as $ax \in E$ or $ax \notin E$ respectively, a contradiction. So, $\mbox{dist}_G(v_2, v_3) = 2$. Similarly, $\mbox{dist}_G(v_4, v_5) = 2$. Hence, there exist $b, d \in V$ such that $bv_2, bv_3, dv_4, dv_5 \in E$. Then $\mbox{dist}_G(v_3, v_4) = 2$ (otherwise, $\{v_2, b, v_3, v_4, c, v_5\}$ will induce either a banner or a $P_6$ in $G$, according as $bd \in E$ or $bd \notin E$ respectively, a contradiction.) So, there exists $c \in V$ such that $cv_3, cv_4 \in E$. Next, we show the following.

\begin{clm}\label{ab-bc-cd-de-in-E} $ab, bc, cd, de \in E$.
\end{clm}
\no{\it Proof of Claim \ref{ab-bc-cd-de-in-E}} : We prove the claim by assuming the contrary as follows:
\begin{enumerate}

\item[(i)] On the contrary, assume that $ab \notin E$. Then $ac \in E$ (otherwise, $\{v_1, a, v_2, b, v_3, $ $c, v_4\}$ will induce a banner or$P_6$ in $G$). Then $bc \notin E$ (otherwise, $G[\{v_1, a, v_2, b,$ $ c\}]$ is a banner in $G$). Then similar to the case of $ac \in E$, we see that $bd \in E$, and hence $ad \notin E$ (otherwise, $G[\{v_1, a, v_2, b, d\}]$ is a banner in $G$). So, $cd \in E$ (otherwise, $G[\{v_1, a, c, v_3, b, d\}]$ is a $P_6$ in $G$). But, then $G[\{v_2, b, v_3, c, d\}]$ is a banner in $G$, a contradiction. So, $ab \in E$. Similarly, $de \in E$.

\item[(ii)]  On the contrary, assume that $bc \notin E$. Then $ac \in E$ (otherwise, since $ab \in E$ (by (i)), we see that $G[\{v_1, a, b, v_3, c, v_4\}]$ is a $P_6$ in $G$). But, then $G[\{v_1, a, b, v_3, c\}]$ is a banner in $G$, a contradiction. So, $bc \in E$. Similarly, $cd \in E$.
  $\blacklozenge$
\end{enumerate}

Then we have the following:

\begin{clm}\label{2-v2-b-v5-d-notin-D} $v_2, b, v_5, d \notin D$.
\end{clm}
\no{\it Proof of Claim \ref{2-v2-b-v5-d-notin-D}} : (i) On the contrary, assume that $v_2 \in D$. Then since $a, v_1 \notin D$, there exists $v_1' \in D$ such that $v_1v_1' \in E$. Also, since $b, v_3, c \notin D$, there exists $v_3' \in D$ such that $v_3v_3' \in E$. Note that by the distance reason, $v_1' \neq v_3'$. Now, $G[\{v_1', v_1, a, b, v_3, v_3'\}]$ is a $P_6$ in $G$, a contradiction. So, $v_2 \notin D$. Similarly, $v_5 \notin D$. (ii) On the contrary, assume that $b \in D$. Then since $a, v_1 \notin D$, there exists $v_1' \in D$ such that $v_1v_1' \in E$. Also, since $c, v_4, d \notin D$, there exists $v_4' \in D$ such that $v_4v_4' \in E$. Note that by the distance reason, $v_1' \neq v_4'$. Now, $\{v_1', v_1, a, b, c, v_4, v_4'\}$ will induce a $P_6$ in $G$, a contradiction. So, $b \notin D$. Similarly, $d \notin D$.   $\blacklozenge$

\vspace{0.25cm}
\noindent{\bf Case 2.1}: Suppose that $bd \notin E$.

Then we have the following:

\begin{clm}\label{2-ac-ce-in-E} $ac, ce \in E$.
\end{clm}
\no{\it Proof of Claim \ref{2-ac-ce-in-E}} : On the contrary, assume that $ac \notin E$. Then $ad \in E$ (otherwise, $G[\{v_1, a, b, c, d, v_5\}]$ is a $P_6$ in $G$). But, then $G[\{v_1, a, b, c, d\}]$ is a banner in $G$, a contradiction. So, $ac \in E$. Similarly, $ce \in E$.
  $\blacklozenge$

\begin{clm}\label{2-a-e-notin-D} $a, e \notin D$.
\end{clm}
\no{\it Proof of Claim \ref{2-a-e-notin-D}} : On the contrary, assume that $a \in D$. Then since $b, v_3, c \notin D$, there exists $v_3' \in D$ such that $v_3v_3' \in E$. Also, since $d, e \notin D$ and $v_5 \notin D$ (by Claim \ref{2-v2-b-v5-d-notin-D}), there exists $v_5' \in D$ such that $v_5v_5' \in E$. Note that by the distance reason, $v_3' \neq v_5'$.  Then since $v_3'c, v_5'c \notin E$ (by the definition of $D$), and since  $v_3'd \notin E$ (else, $G[\{v_3', v_3, c, d, v_5\}]$ is a banner in $G$), we see that $v_5'd \in E$ (otherwise, $G[\{v_3', v_3, c, d, v_5, v_5'\}]$ is a $P_6$ in $G$). Again, since $v_3'e \notin E$ (else, $\{v_3', v_3, c, e, v_6\}$ will induce a banner in $G$), we have $v_5'e \in E$ (otherwise, $G[\{v_3', v_3, c, e, v_5, v_5'\}]$ is a $P_6$ in $G$).
Hence, $v_6 \notin D$, and there exists $v_6' \in D$ such that $v_6v_6' \in E$. Note that $v_6' \neq v_5'$ (otherwise, $G[\{v_3', v_3, c, d, v_5', v_6\}]$ is a $P_6$ in $G$), and by the distance reason, $v_3' \neq v_6'$. But, then $G[\{v_3', v_3, c, e, v_6, v_6'\}]$ is a $P_6$ in $G$, a contradiction. So, $a \notin D$. Similarly, $e \notin D$.   $\blacklozenge$

By Claims \ref{2-v2-b-v5-d-notin-D} and \ref{2-a-e-notin-D}, and by the distance reason, there exist $v_2', v_5' \in D (v_2' \neq v_5')$ such that $v_2v_2', v_5v_5'\in E$. Then we have the following:

\begin{clm}\label{2-v2'b-v5'd-in-E} $v_2'b, v_5'd \in E$.
\end{clm}
\no{\it Proof of Claim \ref{2-v2'b-v5'd-in-E}} : On the contrary, suppose that $v_2'b \notin E$. Then $v_2'c \notin E$ (otherwise, $G[\{v_2', v_2, b, c, v_4\}]$ is a banner in $G$), and hence $v_2'd \in E$ (otherwise, $G[\{v_2', v_2, b, c, d, v_5\}]$ is a $P_6$ in $G$). So, $v_5'd \notin E$. Then by using similar arguments, we deduce that $v_5'c \notin E$ and $v_5'b \in E$. Hence, $v_4, c, d \notin D$, and there exists $v_4' \neq v_2'$ such that $v_4v_4' \in E$. Then we see that $v_4' \neq v_5'$ (otherwise, $G[\{b, v_5', v_4, d, v_5\}]$ is a banner in $G$). But, then $G[\{v_4', v_4, d, v_2', v_2, b\}]$ is a $P_6$ in $G$, a contradiction. So, $v_2'b \in E$. Similarly, $v_5'd \in E$. $\blacklozenge$

Also, we have the following:

\begin{clm}\label{2-v2'a-in-E} $v_2'a, v_5'e \in E$.
\end{clm}
\no{\it Proof of Claim \ref{2-v2'a-in-E}} : Assume the contrary. Then $v_2'c \notin E$ (otherwise, $G[\{v_2', v_2, a, c, $ $ v_4\}]$ is a banner in $G$). Now, if $ad \notin E$, then $G[\{v_2', v_2, a, c, d, v_5\}]$ is a $P_6$ in $G$, a contradiction. So, assume that $ad \in E$. Now, (i) if $v_4v_5' \in E$, then $v_5'c \in E$ (otherwise, $G[\{v_2', b, c, v_4, v_5', v_5\}]$ is a $P_6$ in $G$), and hence $v_5'a \in E$ (otherwise, $G[\{v_2', v_2, a, c, v_5', v_5\}]$ is a $P_6$ in $G$).
Also, $v_2'v_3 \notin E$ (otherwise, $G[\{v_2, v_2', v_3, c, v_5', v_5\}]$ is a $P_6$ in $G$). Thus, since $b, v_3, c \notin D$,there exists $v_3' \neq v_2', v_5'$ such that $v_3v_3' \in E$. But, then $G[\{v_3', v_3, b, a, v_5', v_5\}]$ is a $P_6$ in $G$, a contradiction. So, (ii) assume that $v_4v_5' \notin E$. Since $c, v_4, d \notin D$, there exists $v_4' (\neq v_5')\in D$ such that $v_4v_4' \in E$. Then $v_4'a \notin E$ (otherwise, $G[\{v_4', v_4, d, a, v_1\}]$ is a banner in $G$). But, then $G[\{v_4', v_4, d, a, v_2', v_2\}]$ is a $P_6$ in $G$, a contradiction. Hence, $v_2'a \in E$. By using similar arguments, we can also show that $v_5'e \in E$.  $\blacklozenge$

Since $a, v_1 \notin D$, there exists $v_1' \in D$ such that $v_1v_1' \in E$.

Now, if $v_4v_5' \in E$, then $v_2'c \notin E$ (otherwise, $G[\{v_2, v_2', c, v_4, v_5', v_5\}]$ is a $P_6$ in $G$), and hence $v_5'c \in E$ (otherwise, $G[\{v_2', b, c, v_4, v_5', v_5\}]$ is a $P_6$ in $G$). Then $v_2'v_3 \notin E$ (otherwise, $G[\{v_2, v_2', v_3, c, v_5', v_5\}]$ is a $P_6$ in $G$). Thus, since $b, v_3, c \notin D$, there exists $v_3' \neq v_2', v_5'$ such that $v_3v_3' \in E$. Note that by the distance reason, $v_1' \neq v_3'$, and $v_1' \neq v_2'$ (otherwise, $G[\{v_1, v_2', b, c, v_5', v_5\}]$ is a $P_6$ in $G$). Now, $G[\{v_1', v_1, a, b, v_3, v_3'\}]$ is a $P_6$ in $G$, a contradiction.

So, assume that $v_4v_5' \notin E$. Since $c, v_4, d \notin D$ and by the distance reasons, there exists $v_4' (\neq v_1', v_2', v_5') \in D$ such that $v_4v_4' \in E$.
Next, we prove that $v_1v_2' \notin E$. Suppose not. Then $v_2'c \in E$ (otherwise, $G[\{v_1, v_2', b, c, d, v_5\}]$ is a $P_6$ in $G$). Then $v_5'v_6 \notin E$ (otherwise, $G[\{v_1, v_2', c, d, v_5', v_6\}]$ is a $P_6$ in $G$). Since $e, v_6 \notin D$, there exists $v_6' (\neq v_5', v_4', v_2') \in D$ such that $v_6v_6' \in E$. Then, $G[\{v_1, v_2', c, e, v_6, v_6'\}]$ is a $P_6$ in $G$, a contradiction. So,  $v_1v_2' \notin E$, and hence there exists $v_1' \neq v_2'$. Then $v_1'c \notin E$ (otherwise, $G[\{v_1', v_1, a, c, v_4\}]$ is a banner in $G$), and hence $v_4'c \in E$ (otherwise, $G[\{v_1', v_1, a, c, v_4, v_4'\}]$ is a $P_6$ in $G$). But, then $G[\{v_1', v_1, a, d, v_4, v_4'\}]$ or $G[\{v_1', v_1, a, c, d, v_5\}]$ is a $P_6$ in $G$ according as $ad \in E$ or $ad \notin E$ respectively, a contradiction.

\vspace{0.25cm}
\noindent{\bf Case 2.2}: Suppose that $bd \in E$.

Then either $ad \in E$ or $be \in E$ (otherwise, $G[\{a, b, d, e, v_6\}]$ is a banner in $G$, if $ae \in E$, and $G[\{v_1, a, b, d, e, v_6\}]$ is a $P_6$ in $G$, if $ae \notin E$, a contradiction).  We may assume that $ad \in E$. Then we have the following:

\begin{clm}\label{2-2-a-e-notin-D} $a, e \notin D$.
\end{clm}
\no{\it Proof of Claim \ref{2-2-a-e-notin-D}} : (i) On the contrary, assume that $a \in D$. Then since $b, v_3, c \notin D$, there exists $v_3' \in D$ such that $v_3v_3' \in E$. Also, since $d, v_5, e \notin D$,  there exists $v_5' \in D$ such that $v_5v_5' \in E$. Note that by the distance reason, $v_3' \neq v_5'$. Now, $G[\{v_3', v_3,  b, d, v_5, v_5'\}]$ is a $P_6$ in $G$, a contradiction. So, $a \notin D$. (ii) On the contrary, assume that $e \in D$. Then since $c, v_4, d \notin D$, there exists $v_4' \in D$ such that $v_4v_4' \in E$.  Since $a \notin D$ (by (i)) and $v_2, b \notin D$ (by Claim \ref{2-v2-b-v5-d-notin-D}), there exists $v_2' (\neq v_4')\in D$ such that $v_2v_2' \in E$. Now, if $ae \in E$, then $v_2'a \notin E$, and thus $G[\{v_2', v_2, a, d, v_4, v_4'\}]$ is a $P_6$ in $G$, a contradiction. So, assume that $ae \notin E$. Then $v_2'a \in E$  (otherwise, $G[\{v_2', v_2, a, d, e, v_6\}]$ is a $P_6$ in $G$), and $v_2'b \in E$ (otherwise, $G[\{v_2', v_2, b, d, v_4, v_4'\}]$ is a $P_6$ in $G$). So, $v_1 \notin D$, and there exists $v_1' (\neq v_4') \in D$ such that $v_1v_1' \in E$. Note that $v_1' \neq v_2'$ (otherwise, $G[\{v_1, v_2', b, d, v_4, v_4'\}]$ is a $P_6$ in $G$). But, then $G[\{v_1', v_1, a, d, v_4, v_4'\}]$ is a $P_6$ in $G$, a contradiction. Hence, $e \notin D$. $\blacklozenge$

Then by Claims \ref{2-v2-b-v5-d-notin-D} and \ref{2-2-a-e-notin-D}, there exist $v_2', v_5' \in D (v_2' \neq v_5')$ such that $v_2v_2', v_5v_5' \in E$. Then:

\begin{clm}\label{2-2-v2'b-in-E} $v_2'b \in E$.
\end{clm}
\no{\it Proof of Claim \ref{2-2-v2'b-in-E}} :  Suppose not. Then $v_2'c \notin E$ (else, $G[\{v_2', v_2, b, c,  v_4\}]$ is a banner in $G$). Now, if $v_5'd \notin E$, then since $v_2'd, v_5'b \notin E$ (otherwise, either $\{v_2', v_2, b, d, v_5\}$ or $\{v_5', v_5, b, d, v_2\}$ will induce a banner in $G$), we see that $G[\{v_2', v_2, b, d, v_5, v_5'\}]$ is a $P_6$ in $G$, a contradiction. So, assume that $v_5'd \in E$. Then $v_4v_5' \in E$ (otherwise, since $c, v_4, d \notin D$, there exists $v_4' (\neq v_2', v_5') \in D$ such that $v_4v_4' \in E$. Then since $v_4'b \notin E$ (else, $G[\{v_4', v_4, d, b, v_2\}]$ is a banner in $G$), we have  $G[\{v_2', v_2, b, d, v_4, v_4'\}]$ is a $P_6$ in $G$, a contradiction). Then $v_5'b, v_5'c \in E$ (otherwise, $\{v_2', v_2, b, c, v_4, v_5', v_5\}$ will induce either a $P_6$ or a banner in $G$).

Now, (i) if $be \in E$, then since $v_2'e \notin E$ (else, $G[\{v_2', v_2, b, e, v_6\}]$ is a banner in $G$), we have $v_5'e \in E$ (otherwise, $G[\{v_5', v_5, e, b, v_2\}]$ is a banner in $G$). Then since $v_5'v_6  \notin E$ (by the distance reason), and since $e, v_6 \notin D$, there exists $v_6' (\neq v_2', v_5')\in D$ such that $v_6v_6' \in E$. But, then $G[\{v_6', v_6, e, b, v_2, v_2'\}]$ is a $P_6$ in $G$, a contradiction. So, (ii) assume that $be \notin E$. Then $v_2'e \in E$ (otherwise, $G[\{v_2', v_2, b, d, e, v_6\}]$ is a $P_6$ in $G$). But, then $G[\{v_4, v_5', v_5, e, v_2', v_2\}]$ is a $P_6$ in $G$, a contradiction. Hence the claim holds.
 $\blacklozenge$

Then since $b, c, v_3 \notin D$, there exists $v_3' (\neq v_5') \in D$ such that $v_3v_3' \in E$.

\begin{clm}\label{2-2-v5'd-in-E} $v_5'd \in E$.
\end{clm}
\no{\it Proof of Claim \ref{2-2-v5'd-in-E}} :  Suppose not. Then $v_2'a \in E$ (otherwise, since $v_2'd \notin E$ (else, $G[\{v_2', v_2, a, d,  v_5\}]$ is a banner in $G$), and since $v_5'a \notin E$ (else, $G[\{v_5', v_5, d, a, v_1\}]$ is a banner in $G$), we have $G[\{v_2', v_2, a, d, v_5, v_5'\}]$ is a $P_6$ in $G$, a contradiction).  Now, (i) if $v_1v_2' \in E$, then  since $b, v_3, c \notin D$, there exists $v_3' (\neq v_2', v_5') \in D$ such that $v_3v_3' \in E$. Then $v_3'd  \in E$ (otherwise, $G[\{v_3', v_3, b, d, v_5, v_5'\}]$ is a $P_6$ in $G$). But, then $G[\{v_3', v_3, b, d,  v_5\}]$ is a banner in $G$, a contradiction. (ii) So, assume that $v_1v_2' \notin E$, and hence there exists $v_1' (\neq v_2', v_5') \in D$ such that $v_1v_1' \in E$. Then $v_1'd  \in E$ (otherwise, $G[\{v_1', v_1, a, d, v_5, v_5'\}]$ is a $P_6$ in $G$). But, then $G[\{v_1', v_1, a, d,  v_5\}]$ is a banner in $G$, a contradiction.  $\blacklozenge$

Then since $c, d, v_4 \notin D$, there exists $v_4' (\neq v_2') \in D$ such that $v_4v_4' \in E$.

\begin{clm}\label{2-2-v2'a-in-E} $v_2'a \in E$.
\end{clm}
\no{\it Proof of Claim \ref{2-2-v2'a-in-E}} :  Suppose not.
 Then if $v_2'v_3 \in E$, then  $G[\{v_3, v_2', v_2, a, d,  v_5\}]$ is a $P_6$ in $G$, a contradiction.
 So, assume that $v_2' \neq v_3'$.  Then  $v_5'a \notin E$ (else, $G[\{v_3', v_3, b, a, v_5', v_5\}]$ is a $P_6$ in $G$).
Now, if $v_4' \neq v_5'$, then since $v_4'a \notin E$ (else, $G[\{v_4', v_4, d, a,  v_1\}]$ is a banner in $G$), we have $G[\{v_2', v_2, a, d, v_4, v_4'\}]$ is a $P_6$ in $G$, a contradiction. So, assume that $v_4v_5' \in E$. Then:

(i) If $ae \in E$, then since $v_2'e \notin E$ (else, $G[\{v_2', v_2, a, e,  v_6\}]$ is a banner in $G$), we have $v_5'e \in E$ (otherwise, $G[\{v_2', v_2, a, e, v_5, v_5'\}]$ is a $P_6$ in $G$). But, then $G[\{v_4, v_5', e, a, v_2, v_2'\}]$ is a $P_6$ in $G$, a contradiction.

(ii) If $ae \notin E$, then  since  $v_2'e \in E$ (else, $G[\{v_2', v_2, a, d, e,  v_6\}]$ is a $P_6$ in $G$), we have $be \in E$ (otherwise,$G[\{v_2', b, d, e,  v_6\}]$ is a banner in $G$). But, then $G[\{v_3', v_3, b, e, v_5, v_5'\}]$ is a $P_6$ in $G$, a contradiction.

So the claim holds. $\blacklozenge$

Hence, $v_1 \notin D$, and thus there exists  $v_1v_1'\in E$.

\begin{clm}\label{2-2-v1'-neq-v2'} $v_1'\neq v_2'$.
\end{clm}
\no{\it Proof of Claim \ref{2-2-v1'-neq-v2'}} :  Suppose not. assume that $v_1v_2' \in E$. Then by the distance reason, $v_3' \neq v_2'$. Then:
(i) If $be \in E$, then since $v_3'e \notin E$ (else, $G[\{v_3', v_3, b, e, v_6\}]$ is a banner in $G$), we have $v_5'e \in E$ (else, $G[\{v_3', v_3, b, e, v_5, v_5'\}]$ is a $P_6$), and hence $v_5'v_6 \notin E$ (otherwise, $G[\{v_1, v_2', b, d, v_5', v_6\}]$ is a $P_6$). Since $e, v_6 \notin D$, there exists $v_6' (\neq v_3', v_5') \in D$ such that $v_6v_6' \in E$. But, then $G[\{v_6', v_6, e, b, v_3, v_3'\}]$ is a $P_6$ in $G$, a contradiction. (ii) If $be \notin E$, then $v_2'e \in E$ (else, $G[\{v_1, v_2', b, d, e,  v_6\}]$ is a $P_6$ in $G$). But, then $G[\{v_2', b, d, e,  v_6\}]$ is a banner in $G$, a contradiction. $\blacklozenge$

Then $v_3' = v_2'$ and $v_4' = v_5'$ (otherwise, either $G[\{v_1', v_1, a, b, v_3, v_3'\}]$ or $G[\{v_1',  $ $ v_1,a, d, v_4, v_4'\}]$ is a $P_6$ in $G$). That is, $v_3v_2', v_4v_5' \in E$. Then: (i) If $ae \in E$, then since $v_1'e \notin E$ (else, $G[\{v_1', v_1, a, e,  v_6\}]$ is a banner in $G$), we have  $v_5'e \in E$ (otherwise, $G[\{v_1', v_1, a, e, v_5, v_5'\}]$ is a $P_6$ in $G$). But, then $G[\{v_1', v_1, a, e, v_5', v_4\}]$ is a $P_6$, a contradiction. (ii) If $ae \notin E$, then $v_2'e \in E$ (else, $G[\{v_3, v_2', a, d, e,  v_6\}]$ is a $P_6$ in $G$). But, then $G[\{v_4, v_5', v_5, e, v_2', v_2\}]$ is a $P_6$ in $G$, a contradiction.

Since the other cases are symmetric, we have proved the theorem. \hfill{$\Box$}

Next, we prove the following:

\begin{thm}\label{P6-banner-free-ed-implies-banner-free}Let $G = (V, E)$ be a ($P_6$, banner)-free graph. If $G$ has an efficient dominating set, then $G^2$ is banner-free.
\end{thm}

\no{\it Proof of Theorem \ref{P6-banner-free-ed-implies-banner-free}} : Let $G$ be a ($P_6$, banner)-free graph having an efficient dominating set $D$, and assume to the contrary that $G^2$ contains an induced
banner, say with vertices $v_1, v_2, v_3, v_4,$ and $v_5$,  and edges $v_1v_2, v_2v_3, v_3v_4, v_4v_1,$ and $v_3v_5$.  Then $\mbox{dist}_G(v_1, v_2) \leq 2, \mbox{dist}_G(v_2, v_3) \leq 2, \mbox{dist}_G(v_3, v_4) \leq 2$, $\mbox{dist}_G(v_4, v_1) \leq 2$, and  $\mbox{dist}_G(v_3, v_5) \leq 2$, while $\mbox{dist}_G(v_1, v_3) \geq 3, \mbox{dist}_G(v_1, v_5) \geq 3, \mbox{dist}_G(v_2, v_4) \geq 3, \mbox{dist}_G(v_2, v_5) \geq 3$, and $\mbox{dist}_G(v_4, v_5) \geq 3$.  We often use these distance properties implicitly in the remaining proof.

\vspace{0.25cm}

\no{\bf Case 1}: Suppose that $\mbox{dist}_G(v_3, v_5) = 1$.

Then since $\mbox{dist}_G(v_2, v_5) \geq 3$, $\mbox{dist}_G(v_2, v_3) = 2$. Again, since $\mbox{dist}_G(v_4, v_5) \geq 3$, we have $\mbox{dist}_G(v_3, v_4) = 2$. So, there exist vertices $a$ and $b$ in $V$ such that $av_2, av_3, bv_3, bv_4 \in E$.  Since $\mbox{dist}_G(v_2, v_4) \geq 3$, at least one of  $\mbox{dist}_G(v_1, v_4)$, $\mbox{dist}_G(v_1, v_2)$ is equal to two. We may assume (wlog.) that $\mbox{dist}_G(v_1, v_2) = 2$. Hence, there exists $c \in V$ such that $cv_1, cv_2 \in E$. Then $ac\in E$ (otherwise, $\{v_5, v_3, a, v_2, c, v_1\}$ will induce a $P_6$ in $G$).
Then $\mbox{dist}_G(v_1, v_4) = 2$ (otherwise, $\{v_5, v_3, a, c, v_1, v_4\}$ will induce a $P_6$ in $G$). Hence, there exists $d \in V$ such that $dv_1, dv_4 \in E$. As earlier, $bd \in E$, and hence $cd \in E$ (otherwise, either $G[\{v_5, v_3, b, d, v_1, c\}]$ is a $P_6$ in $G$ or $G[\{v_3, b, d, v_1, c\}]$ is a banner in $G$ according as $bc \notin E$ or $bc \in E$ respectively, a contradiction). Then $bc, ad \in E$ (otherwise, either $G[\{v_5, v_3, b, d, c, v_2\}]$ is a $P_6$ in $G$ or $G[\{v_5, v_3, a, c, d, v_4\}]$ is a $P_6$ in $G$), and hence $ab \in E$ (otherwise, $G[\{v_5, v_3, b, d, a\}]$ is a banner in $G$). Then we have the following:

\begin{clm}\label{21-v2-v4-a-b-c-d-notin-D}$v_2, v_4, a, b, c, d \notin D$.
\end{clm}

\no{\it Proof of Claim \ref{21-v2-v4-a-b-c-d-notin-D}} : We prove the claim by assuming the contrary one by one as follows:
\begin{enumerate}

\item[(i)] On the contrary, assume that $v_2 \in D$. Then by the definition of $D$, we have $d, c, v_1 \notin D$. So, there exists $v_1' \in D$ such that $v_1v_1' \in E$. Then by the distance properties and by the definition of $D$, we see that $G[\{v_1', v_1, c, a, v_3, v_5\}]$ is a $P_6$ in $G$, a contradiction. Hence, $v_2 \notin D$. Similarly, $v_4 \notin D$.

\item[(ii)] On the contrary, assume that $a \in D$. Then by the definition of $D$, we have $d, b, v_4 \notin D$. So, there exists $v_4' \in D$ such that $v_4v_4' \in E$. Then by the definition of $D$ and by using the distance properties, we see that $G[\{v_4', v_4, d, a, v_3, v_5\}]$ is a $P_6$ in $G$, a contradiction. Hence, $a \notin D$. Similarly, $b \notin D$.

\item[(iii)] On the contrary, assume that $c \in D$. Then since $b, d, v_4 \notin D$, there $v_4' \in D$ such that $v_4v_4' \in E$.
   Then by distance reasons and by the definition of $D$, we have $v_4'v_3 \in E$ (otherwise, $G[\{v_4', v_4, d, a, v_3, v_5\}]$ is a $P_6$ in $G$), and hence  $G[\{v_4', v_4, b, v_3, c\}]$ is a banner in $G$, a contradiction. Hence, $c \notin D$. Similarly, $d \notin D$. $\blacklozenge$
   \end{enumerate}

So, there exist $v_2', v_4' \in D$ such that $v_2v_2', v_4v_4' \in E$. Note that by the distance reason, we have  $v_2' \neq v_4'$. Now, we prove the following:

\begin{clm}\label{21-v2'd-v4'c-in-E}$v_2'c, v_4'd \in E$.
\end{clm}

\no{\it Proof of Claim \ref{21-v2'd-v4'c-in-E}} : On the contrary, assume that $v_2'c \notin E$. Then $v_2'b \notin E$ (else, $G[\{v_2', v_2, c, b, v_4\}]$ is a banner in $G$), and $v_2'v_3 \in E$ (else, $G[\{v_2', v_2, c, b, v_3, v_5\}]$ is a $P_6$ in $G$). But, then $G[\{v_5, v_3, v_2', v_2, c, v_1\}]$ is a $P_6$ in $G$, a contradiction. So, $v_2'c \in E$. Similarly, $v_4'd \in E$.  $\blacklozenge$

So, $v_1 \notin D$, and hence there exists $v_1' \in D$ such that $v_1v_1' \in E$. Then, we show the following:

\begin{clm}\label{21-v1'-neq-v2'-v4'} $v_1' \neq v_2', v_4'$.
\end{clm}

\no{\it Proof of Claim \ref{21-v1'-neq-v2'-v4'}} : Assume the contrary, and assume that $v_1' = v_2'$. That is, $v_1v_2' \in E$.  Then since $v_2'v_3 \notin E$ (by the distance reason), we have $v_2'b \in E$ (otherwise, $G[\{v_5, v_3, b, d, v_1, v_2'\}]$ is a $P_6$ in $G$). But, then $G[\{v_2', v_1, d, b, v_4'\}]$ is a banner in $G$, a contradiction. So, $v_1' \neq v_2'$. Similarly, $v_1' \neq v_4'$. $\blacklozenge$

Now, since $v_1'c \notin E$ (by the definition of $D$) and since $v_1'a \notin E$ (else, $G[\{v_1', v_1, c, a, v_3\}]$ is a banner in $G$), we see that
by the distance properties, $G[\{v_1', v_1, c, a, v_3, v_5\}]$ is a $P_6$ in $G$, a contradiction.

\vspace{0.25cm}
\no{\bf Case 2}: Suppose that $\mbox{dist}_G(v_3, v_5) = 2$.

So, there exists $a \in V$ such that $av_3, av_5 \in E$. Then $\mbox{dist}_G(v_2, v_3) = 2 = \mbox{dist}_G(v_3, v_4)$ (otherwise, if  $\mbox{dist}_G(v_2, v_3) = 1$, then  $\mbox{dist}_G(v_1, v_2) = 1$ (else, there exists $x \in V$ such that $xv_1, xv_2 \in E$, and hence $\{v_5, a, v_3, v_2, x, v_1\}$ will induce either a $P_6$ or a banner in $G$, a contradiction). Hence, a contradiction to the fact that $\mbox{dist}_G(v_1, v_3) \geq 3$. A similar contradiction can be arrived if $\mbox{dist}_G(v_3, v_4) = 1$). So, there exist $b, c \in V$ such that $bv_2, bv_3, cv_3, cv_4 \in E$.

Since $\mbox{dist}_G(v_2, v_4) \geq 3$, we may assume that $\mbox{dist}_G(v_1, v_4) = 2$, and  there exists $d \in V$ such that $dv_1, dv_4 \in E$. Then $\mbox{dist}_G(v_1, v_2) = 2$ (otherwise, $\{v_4, d, v_1, v_2, $ $ b, v_3\}$ will induce either a $P_6$ or a banner in $G$), and hence there exists $e \in V$ such that $ev_1, ev_2 \in E$.

\begin{clm}\label{22-edges-in H} $ac, ab, bc, cd, de, be\in E$.
\end{clm}

\no{\it Proof of Claim \ref{22-edges-in H}} : (i) On the contrary, assume that $ac \notin E$. Then $ad \in E$ (otherwise, $\{v_5, a, v_3, c, v_4, d,$ $ v_1\}$ will induce a $P_6$ in $G$), and hence $cd \notin E$ (otherwise, $G[\{v_5, a, v_3, c, d\}]$ is a banner in $G$). Now,
if $ab \notin E$, then $bd \notin E$ (else, $G[\{v_5, a, v_3, b, d\}]$ is a banner in $G$). Then $cb \in E$ (otherwise, $G[\{v_1, d, v_4, c, v_3, b\}]$ is a $P_6$ in $G$). But, then $G[\{v_5, a, d, v_4, c, b\}]$ is a $P_6$ in $G$, a contradiction.
So, assume that $ab \in E$, then $be \in E$ (otherwise, $\{v_5, a, b, v_2, e, v_1\}$ will induce either a $P_6$ or banner in $G$). Then $bc \in E$ (otherwise, $\{v_1, e, b, v_3, c, v_4\}$ will induce either a $P_6$ or banner in $G$), and hence $bd \notin E$ (otherwise, $G[\{d, v_4, c, b, v_2\}]$ is a banner in $G$). But, then $G[\{v_1, d, v_4, c, b, v_2\}]$ is a $P_6$ in $G$, a contradiction. So, $ac \in E$. Similarly, $ab \in E$.

(ii)  On the contrary, assume that $cd \notin E$. Then, $\{v_5, a, c, v_4, d, v_1\}$ will induce either a $P_6$ or banner in $G$, a contradiction. So, $cd \in E$. Similarly, $be \in E$.

(iii)  On the contrary, assume that $bc \notin E$. Then, $\{v_4, c, v_3, b, e, v_1\}$ will induce either a $P_6$ or banner in $G$, a contradiction. So, $bc \in E$. Again, by using similar arguments, we see that $de \in E$.  $\blacklozenge$

Next, we show the following:

\begin{clm}\label{22-vertices-notin-D} $v_2, v_4, b, c, d, e \notin D$.
\end{clm}

\no{\it Proof of Claim \ref{22-vertices-notin-D}} : We prove the claim by assuming the contrary one by one as follows:
\begin{enumerate}

\item[(i)] On the contrary, assume that $v_2 \in D$. Then since $d, e, v_1 \notin D$, there exists $v_1' \in D$ such that $v_1v_1' \in E$. Also,  since $a, b, c, v_3 \notin D$, there exists $v_3' \in D$ such that $v_3v_3' \in E$. By the distance reason, $v_1' \neq v_3'$. But, then  $G[\{v_1', v_1, e, b, v_3, v_3'\}]$ is a $P_6$ in $G$, a contradiction. Hence, $v_2 \notin D$. Similarly, $v_4 \notin D$.

\item[(ii)] On the contrary, assume that $c \in D$. Then since $a, v_5 \notin D$, there exists $v_5' \in D$ such that $v_5v_5' \in E$. Also,  since $e, b, v_2 \notin D$, there exists $v_2' \in D$ such that $v_2v_2' \in E$. By the distance reason, $v_2' \neq v_5'$. But, then $G[\{v_5', v_5, a, b, v_2, v_2'\}]$ is a $P_6$ in $G$, a contradiction. Hence, $c \notin D$. Similarly, $b \notin D$.

\item[(iii)] On the contrary, assume that $d \in D$. Then since $e, b, v_2 \notin D$, there exists $v_2' \in D$ such that $v_2v_2' \in E$. Also,  since $a, b, c,  v_3 \notin D$, there exists $v_3' \in D$ such that $v_3v_3' \in E$. Then, $v_2' \neq v_3'$ (otherwise, $G[\{v_2, v_2' (= v_3'), v_3, c, d, v_1\}]$ is a $P_6$ in $G$). But, then $\{v_3', v_3, c, d, e, v_2, v_2'\}$ will induce a  $P_6$ in $G$, a contradiction. Hence, $d \notin D$. Similarly, $e \notin D$.

   \end{enumerate}

\no{} By (i), (ii) and (iii), we see that Claim \ref{22-vertices-notin-D} is proved.  $\blacklozenge$

So, there exist $v_2', v_4' \in D$ such that $v_2v_2', v_4v_4' \in E$. Note that by the distance reason, $v_2' \neq v_4'$.

\begin{clm}\label{22-v1-notin-D} $v_1 \notin D$.
\end{clm}

\no{\it Proof of Claim \ref{22-v1-notin-D}} : If not, then $G[\{v_4', v_4, d, e, v_2, v_2'\}]$ is a $P_6$ in $G$, a contradiction. Hence, $v_1 \notin D$. $\blacklozenge$

Since $d, e, v_1 \notin D$ (by Claims \ref{22-vertices-notin-D} and \ref{22-v1-notin-D}), there exists $v_1' \in D$ such that $v_1v_1' \in E$. Then:

\begin{clm}\label{22-v1'-neq-v2'-v4'} $v_1' \neq v_2'$ and $v_1' \neq v_4'$.
\end{clm}

\no{\it Proof of Claim \ref{22-v1'-neq-v2'-v4'}} : On the contrary, suppose that $v_1' = v_2'$. That is, $v_1v_2' \in E$. Then either $v_4'd \in E$ or $v_2'd \in E$ (not both) (otherwise, $G[\{v_4', v_4, d, v_1, v_2', v_2\}]$ is a $P_6$ in $G$). Now, if $v_4'd \in E$ and $v_2'd \notin E$, then by using the distance properties, we see that $v_2'c \in E$ (otherwise, $G[\{v_3, c, d, v_1, v_2', v_2\}]$ is a $P_6$ in $G$). But, then $G[\{v_1, v_2', c, d, v_4'\}]$ is a banner in $G$, a contradiction. So, assume that $v_2'd \in E$ and $v_4'd \notin E$. Then $v_2'e \in E$ (otherwise, $G[\{e, v_2, v_2', d, v_4\}]$ is a banner in $G$).  Then $v_2'b \in E$ (otherwise, since $bd \in E$ (else, $G[\{v_4, d, v_2', v_2, b, v_3\}]$ is a $P_6$ in $G$), we see that $G[\{v_4, d, b, v_2, v_2'\}]$ is a banner in $G$). Now, if $v_4'v_3 \in E$, then $G[\{v_4, v_4', v_3, b, v_2', v_1\}]$ is a $P_6$ in $G$, a contradiction. So, assume that $v_4'v_3 \notin E$. Since $v_3 \notin D$, there exists $v_3' (\neq v_2', v_4') \in D$ such that $v_3v_3' \in E$. Now, $\{v_3', v_3, b, e, d, v_4, v_4'\}]$ will induce a $P_6$ in $G$, a contradiction. Hence the claim holds. $\blacklozenge$

Then either $v_4'd \in E$ or $v_2'e \in E$ (otherwise, since $G[\{v_4', v_4, d, e, v_2, v_2'\}]$ is not a $P_6$ in $G$, either $v_4'e \in E$ or $v_2'd \in E$. Then either $G[\{v_4, v_4', d, e, v_2\}]$ or $G[\{v_2, v_2', d, e, v_4\}]$ is a banner in $G$, a contradiction).  We may assume that $v_4'd \in E$. Then $v_1'c \notin E$ (otherwise, $G[\{v_1', v_1, d, c, v_3\}]$ is a banner in $G$).

Now, if $ad \notin E$, then $v_1'a \in E$ (otherwise, $G[\{v_1', v_1, d, c, a, v_5\}]$ is a $P_6$ in $G$). But, then $G[\{v_5, a, v_1', v_1,$ $ d,  v_4'\}]$ is a $P_6$ in $G$, a contradiction.

So, assume that $ad \in E$. Then $v_1'a \notin E$ (else, $G[\{v_1', v_1, d ,a, v_5\}]$ is a banner in $G$). Now, we prove the following.

\begin{clm}\label{22-bd-in-E} $bd \in E$.
\end{clm}

\no{\it Proof of Claim \ref{22-bd-in-E}} : If not, then $v_1'b \in E$ (otherwise, $G[\{v_1', v_1, d, c, b, v_2\}]$ is a $P_6$ in $G$). But, then
$G[\{d, v_1, v_1', b, v_2, v_2'\}]$ is a $P_6$ in $G$, a contradiction. Hence,  $bd \in E$. $\blacklozenge$

Then since $v_1'b \notin E$ (otherwise, $G[\{v_1', v_1, d, b, v_3\}]$ is a banner in $G$), we have $v_2'b \in E$ (otherwise, $G[\{v_1', v_1, d, b, v_2, v_2'\}]$ is a $P_6$ in $G$). So, since $a, b, c, v_3 \notin D$, there exists $v_3' (\neq v_1') \in D$ such that $v_3v_3' \in E$. Then:

\begin{clm}\label{22-v3'-neq-v2'-v4'} $v_3' \neq v_2', v_4'$. That is, $v_3v_2', v_3v_4' \notin E$.
\end{clm}

\no{\it Proof of Claim \ref{22-v3'-neq-v2'-v4'}} : Suppose not. If $v_3v_2' \in E$, then $\{v_1', v_1, d, a, v_3, v_2', v_2\}$ will induce a $P_6$ in $G$, and if $v_3v_4' \in E$, then $G[\{v_3, v_4', d, b, v_2\}]$ is a banner in $G$, a contradiction. So, the claim holds. $\blacklozenge$

Hence, $G[\{v_1', v_1, d, b, v_3, v_3'\}]$ is a $P_6$ in $G$, a contradiction. \hfill{$\Box$}

\begin{thm}\label{EDS-P6-banner-free}
 The EDS can be solved in polynomial time for ($P_6$, banner)-free graphs.
 \end{thm}

 \no{\it Proof of Theorem \ref{EDS-P6-banner-free}} : Since the MWIS problem in ($P_6$, banner)-free graphs can be solved in polynomial time \cite{BKLM, K}, the theorem follows by
 Theorems \ref{P6-banner-free-ed-implies-P6-free} and \ref{P6-banner-free-ed-implies-banner-free}, and Lemma \ref{EDS-MWIS-reduction}. \hfill{$\Box$}

\end{document}